\documentclass[twocolumn,showpacs,prb]{revtex4}
\usepackage{graphicx}
\usepackage{dcolumn}
\usepackage{amsmath}
\usepackage{latexsym}
\usepackage{longtable}
\begin{document}
\title{Electrical transport properties of nanostructured ferromagnetic perovskite oxides $\mathrm{La_{0.67}Ca_{0.33}MnO_3}$ and $\mathrm{La_{0.5}Sr_{0.5}CoO_3}$ at low temperatures (5 K $\geq T \geq 0.3$ K) and high magnetic field} 
    
\author{Tapati Sarkar\footnote[1]{Present address: Laboratoire CRISMAT, UMR 6508 CNRS ENSICAEN, 6 bd Marechal Juin, 14050 CAEN, France. Email: tapati.sarkar@ensicaen.fr}, M. Venkata Kamalakar\footnote[2]{Present address: IPCMS, Department of Magnetic Objects on the Nanoscale, 23 rue du Loess BP 43, F-67034, Strasbourg Cedex 2, France. Email: venkata@ipcms.u-strasbg.fr} and  A. K. Raychaudhuri\footnote[3]{email:arup@bose res.in}}
\affiliation{DST Unit for NanoSciences, Department of Condensed Matter Physics and Materials Science,S.N.Bose National Centre for Basic Sciences, Block JD, Sector III, Salt Lake, Kolkata 700 098, West Bengal, India.}

\date{\today}
\begin{abstract}

We report a comprehensive study of the electrical and magneto-transport properties of nanocrystals of $\mathrm{La_{0.67}Ca_{0.33}MnO_3}$ (LCMO) (with size down to 15 nm) and $\mathrm{La_{0.5}Sr_{0.5}CoO_3}$ (LSCO) (with size down to 35 nm) in the temperature range 0.3 K to 5 K and magnetic fields upto 14 T. The transport, magnetotransport and non-linear conduction (I-V curves) were  analysed  using the concept of Spin Polarized Tunnelling in the presence of Coulomb blockade. The activation energy of transport, $\Delta$, was used to estimate the tunnelling distances and the inverse decay length of the tunnelling wave function ($\chi$) and the height of the tunnelling barrier ($\Phi_{B}$). The magnetotransport data were used to find out the magnetic field dependences of these tunnelling parameters. The data taken over a large magnetic field range allowed us to separate out the MR contributions at low temperatures arising from tunnelling into two distinct contributions. In LCMO, at low magnetic field, the transport and the MR are dominated by the spin polarization, while at higher magnetic field the MR arises from the lowering of the tunnel barrier by the magnetic field leading to an MR that does not saturate even at 14 T. In contrast, in LSCO, which does not have substantial spin polarization, the first contribution at low field is absent, while the second contribution related to the barrier height persists. The idea of inter-grain tunnelling has been validated by direct measurements of the non-linear I-V data in this temperature range and the I-V data was found to be strongly dependent on magnetic field. We made the important observation that a gap like feature (with magnitude $\sim E_C$, the Coulomb charging energy) shows up in the conductance g(V) at low bias  for the systems with smallest nanocrystal size at lowest temperatures (T $\leq$ 0.7 K). The gap closes as the magnetic field and the temperature are increased. 

\end{abstract}
\pacs{73.63.-b, 72.25.-b}
\maketitle

\noindent
\section{\bf INTRODUCTION}
Electrical conduction in nanostructured materials consisting of granular metals dispersed in insulating oxide matrix has been investigated since the 1970s\cite{Abeles1}. The transport from one metallic grain to the other takes place through the intervening insulator (which may act as the tunnelling barrier) and the transport is governed by the Coulomb charging energy $E_{C}$ of the metallic  grains. Generally at temperatures T $< \frac {E_C}{k_B}$ the current through such a system is limited by the phenomenon of Coulomb blockade (CB) that hinders the current for bias V $< E_{C}$. Often these systems show an activated transport with the resistivity showing the relation\cite{Abeles1}:
\begin{eqnarray}
\rho = \rho_{0}exp(2\sqrt{\frac{\Delta}{k_BT}})
\end{eqnarray}
where, $\rho_{0}$ is a constant inversely proportional to the strength of hopping between sites, and $\Delta$, the activation energy, is  related to the ratio $\frac{s}{d}$, where $\mathrm{s}$ is the average tunnelling distance of the carrier ($\sim$ intergrain separation) and $\mathrm{d}$ is the average grain diameter. The activation energy, $\Delta$, as discussed later on, is related to the charging energy $E_C$. This phenomenon becomes even more fascinating when the metallic grains are ferromagnetic. In such granular magnetic medium, the magnetic energy plays a role and the transport gets modified by the spin order\cite{Abeles2}.  In such cases, the tunnelling between metallic grains depends on the spin polarization of the tunnelling electrons and the magnetization of the ferromagnetic metallic grains. Application of the magnetic field, thus, can change this spin polarized tunnelling (SPT) which gives rise to a change in the current through such systems, leading to the phenomenon of tunnelling magneto resistance (TMR)\cite{julliere,moodera,inoue}. In arrays of magnetic nanocrystals separated by an insulating tunnelling barrier, one would expect the phenomenon of SPT to be strongly modulated by the contribution from Coulomb charging at temperatures T$ < E_C/k_B$. The interplay of these two phenomena leads to interesting physics where the transport in the Coulomb blockade region can be effectively controlled by an applied magnetic field\cite{Yakushijia}. Most of the reported investigations of the interplay of SPT and CB in granular medium have been done in systems containing nanoparticles of conventional metallic ferromagnets like Co and Ni embedded in insulating matrix made of SiO$_2$ or Al$_2$O$_3$\cite{Yakushijia}. In this prototypical system, there is a clear separation of the insulating matrix and the dispersed phase of metallic nanoparticles. In recent years, other systems where the insulating layer can be a non-oxide, have been investigated. For example, in arrays of  CoFe magnetic nanoparticles  where the nanoparticles have been separated by a thin organic insulating layer have been studied to understand the transport in the CB regime\cite{Tan}. In this context, systems made up of ferromagnetic metallic oxide nanoparticles, particularly those of perovskite manganites and cobaltates, present an interesting class of systems. In this case, the barrier need not be another material. The grain boundaries in some of these oxides (in particular the manganites) can be highly resistive and can act as a tunnelling barrier, in particular when they are cooled to lower temperatures. One would expect that in such a system, for sufficiently small crystal size and at sufficiently low temperatures ($k_BT < E_C$), the Coulomb charging effects can show up and one will be able to observe that in I-V measurements. If the ferromagnetic nanocrystals have sufficient spin polarization, then one would expect contribution from SPT as the electron tunnels between neighbouring ferromagnetic nanocrystals.

\noindent
In this paper, we report an investigation of the interplay of SPT and CB by measurement of electronic transport (including non-linear transport (I-V)) in samples of very well characterized $\mathrm {La_{0.67}Ca_{0.33}MnO_{3}}$ (LCMO) nanocrystals with  size down to 15 nm and  temperatures below 5 K and particularly down to 0.3 K, and in magnetic fields upto 14 T. (Note: $\mathrm {La_{0.67}Ca_{0.33}MnO_{3}}$ having particle size in the nanometer regime were synthesized as early as 1996\cite{firstreport}.) This system is known to have highly resistive grain boundaries, which can form the tunnelling barrier. For comparison, we have also carried out similar experiments in nanocrystals of  another perovskite ferromagnetic oxide $\mathrm{La_{0.5}Sr_{0.5}CoO_3}$ with size as small as 35 nm. In this system, the resistance is not very high and the degree of spin polarization is very low. This is expected to show weak manifestations of the above effects. A comparison of the two brings out the special nature of the manganite system and the important role the grain boundaries can play. Investigations on transport in nanocrystals of manganites like LCMO have been done before, however, at temperatures above 5 K.  Below we give a brief review of some of the relevant published works on this problem to put the present investigation in proper perspective and to high light the new observations that are only seen at lower temperatures. 

\noindent
It had been recognized since long that the grain boundary dominated transport in fine particles of manganites can enhance their magnetoresistance\cite{Mahesh} at low temperature, particularly at low fields. In subsequent investigations this was attributed to the SPT effects~\cite{Huang,Arunava}. Presence of Coulomb Blockade had been first investigated in fine powders of $La_{0.67}Sr_{0.33}MnO_{3}$ (LSMO) with grain size down to 20 nm for temperatures down to 10 K\cite{Balcells}. The observed temperature dependence of resistivity was found to obey Eqn. 1 although the activation energy $\Delta$ was identified with $E_{C}$. No non-linear transport (I-V curves) and effect of magnetic field on the CB phenomenon was investigated. Subsequent studies down to 4 K done on $La_{0.67}Ca_{0.33}MnO_{3}$ with grain sizes down to 12 nm addressed the issue of grain size distribution, leading to a distribution of charging energies\cite{garcia}. It was found that the activation energy, $\Delta$, decreases with an increase of the particle diameter $\mathrm{d}$, although the exact dependence was not explored and tunnelling parameters were not evaluated. It was proposed that as the field increases, the coupling between grains is enhanced, leading to de-localization of the charges to neighbouring grains. This reduces the effective capacitance and thus the charging energy. Spin polarized transport (in presence of Coulomb charging) in thin granular films of LCMO were investigated to explore the role of magnetic energy through temperature and field dependent transport down to 30 K. The resistivity was found to obey Eqn. 1, although there was no correlation established between any dimension / size with the activation energy and no I-V characteristics were studied\cite{zeiss}. Electrical transport in granular LCMO with somewhat larger particle size ($d \sim$ 300 nm) had been investigated\cite{Yuan} and modelled, although the resistivity does not show any characteristic upturn at low temperatures. The issue of spin polarized transport has also been investigated\cite{Dey} in nanocrystals of LSMO and LCMO close to room temperature through studies of size dependence of Low field magnetoresistance (LFMR) and a dependence of the LFMR and surface spin susceptibility was established. Due to high temperature the issue of CB was not addressed.

\noindent
The tunnelling nature of transport through grain boundary (GB) junctions using I-V characteristics was investigated in artificial grain boundary junctions in epitaxial films down to 4 K and in magnetic field\cite{Hofener,Mandar}. Although the issue of SPT through the artificial GB junction was investigated, the Coulomb charging was of no consequence in such junctions. Investigation of I-V characteristics of nanocrystals of LCMO and LSMO with size down to 30 nm  were carried out for temperatures  down to 4 K and in magnetic fields upto 9 T  and the results were analysed using the theories of barrier tunnelling\cite{Nieb}. This investigation brought into picture the important estimate of the barrier to tunnelling occurring at the grain boundaries. The SPT in nanostructured films (grain sizes $\sim$ 30 nm) of LCMO and LSMO had been probed through investigations of I-V curves and magnetotransport in magnetic fields upto 6 T and down to 4 K\cite{barnali,shantha}. The films showed prominent rise in resistivity below 40 K, although there was no clear signature of Coulomb charging. The magnetic field had a big effect on the non-linear transport, as seen through the modification of the dynamic conductance in magnetic field. Very recently the low temperature transport including non-linear transport in different nanostructured manganites (including nanostructured films) with grain sizes down to 50 nm have been investigated\cite{Mukhopadhyay} in the temperature range down to 3 K and in magnetic fields upto 10 T. The data were analysed using a description of spin-dependent transport incorporating both tunnelling and non-tunnelling transport. It has been argued that the magnetic field dependent non-linear transport in such systems (with larger particle size $d >$ 50 nm) need not arise from CB phenomenon alone.

\noindent
A review of the published papers on electronic transport in nanostructured manganites shows that while the issue of SPT as well as CB have been raised in earlier investigations in the manganite systems, the effect of both the effects occurring together\cite{Quintela} has not been studied in details. In particular, their contributions to non-linear I-V curves (and its link to the resistivity $\rho$) as well as the effect of high magnetic field have not been addressed or unambiguously understood. There was no observation of a gap like feature in bias dependent conductance curve that shows existence of a charging energy $E_{C}$, the main reason being that the experiments had not been done to low enough temperature (T$< \frac{E_C}{k_B}$) with small enough particles and high magnetic fields to clearly identify various effects that may govern the transport in such nanostructured materials. Often the low temperature rise in resistivity in such systems is taken as a sign of entry to the CB dominated regime although it was realized  that such a rise in $\rho$ at low T does not necessarily imply existence  CB\cite{Andres}. In most of these studies, the size dependence of the observed $\Delta$, the tunnelling parameters and the effect of the magnetic field on these  important parameters have not been investigated. In particular, none of the studies made a quantitative analysis of the activation energy of the transport in the CB regime using the concept of size dependence that can clearly give the average tunnelling distance $\mathrm{s}$, the inverse decay length of the tunnelling wave function $\chi$ (and thus the barrier height)\cite{Abeles1} and also their field dependence if any. We also find that no studies on magnetotransport and I-V characteristics of the nanostructured cobaltate system $\mathrm{La_{0.5}Sr_{0.5}CoO_3}$ (LSCO) have been made except a preliminary report by our group\cite{TapatiJNN}.
 
\noindent
The present investigation was undertaken to carry out these experiments to temperatures as low as 0.3 K and magnetic fields upto 14 T and to quantitatively analyse  the transport ($\rho$ vs T), magnetoresistance (MR) and the non-linear I-V data in these nanostructured materials. In this temperature range, for most of the nanocrystals of LCMO, the Coulomb charging energy $E_{C}$ is larger than $k_{B}T$. We find that a distinct Coulomb blockade dominated transport regime develops below 5 K. The transport was found to contain contributions of transport paths that are mostly of tunnelling type. $\Delta$, obtained from the temperature dependence, was found to depend on the ratio $\frac{s}{d}$, where $\mathrm{d}$ is the average diameter of the particle and $\mathrm{s}$, the average inter-particle separation. This dependence also allowed us to determine the inverse decay length of the tunnelling wave function $\chi$ and the height of the tunnelling barrier. At low magnetic fields, the transport and the MR are dominated by the spin polarized tunnelling, while at higher magnetic fields, the MR arises from the lowering of the tunnel barrier by the magnetic field leading to an MR that does not saturate even at 14 T.  The non linear I-V data were analysed and connection to parameters obtained from the resistivity data was established. We observed that the bias dependence of the non-linearity shows a conventional barrier dominated type tunnelling which interestingly becomes rather large at high magnetic fields. The investigation on LSCO allows us a comparison with the manganite system and elucidates the role of high spin-polarization and  the nature of grain boundary in the manganite systems.
  
\section{EXPERIMENTAL}
The present investigation was done on two well characterized nanostructured oxides, namely, $La_{0.67}Ca_{0.33}MnO_{3}$ (LCMO) nanocrystals with size down to 15 nm and nanocrystals of $\mathrm{La_{0.5}Sr_{0.5}CoO_3}$ (LSCO) with size down to 35 nm. The ferromagnetic state in LCMO with size down to 15 nm has been investigated in details by us recently, using magnetization as well as neutron scattering\cite{NJP}. It has been found that even down to this size range the long range ferromagnetic order prevails in LCMO with $T_{C} \approx 235-240$ K, albeit with a spontaneous magnetization (measured from Neutron scattering data) which is 50 $\%$ of the bulk data. The ferromagnetic order also persists in the $\mathrm{La_{0.5}Sr_{0.5}CoO_3}$ nanocrystals down to the size range investigated.

\noindent
The samples were synthesized by a polymeric (polyol) precursor route to synthesize the nanocrystals\cite{shanthaJMR}. This method allows synthesis at a significantly lower sintering temperature compared to the conventional solid state procedure.  Typical phase formation occurs at $\approx$ $650^0$C for LCMO and $\approx$ $800^0$C for LSCO. Since the phase formation occurs in LSCO at higher temperatures, the nanocrystals grow to a relatively larger size compared to that in LCMO.  We optimized these process parameters to obtain phase pure LCMO and LSCO nanocrystals which were tested through a number of characterization techniques. The details of the process as well as the characterization used have been given in our earlier publications on nanomanganites\cite{NJP,PRB}.
 
\noindent
The  stoichiometry  of  the  materials  were  also checked independently by a quantitative analysis  using  Inductively  Coupled  Plasma  Atomic  Emission  Spectroscopy (ICPAES) and also iodometric titration. All our samples have oxygen stoichiometry close to the ideal stoichiometry with a slight oxygen deficiency $\approx 0.02$. The nanocrystals were also characterized using powder X-Ray  Diffraction (XRD) and high resolution Transmission electron Microscope (HRTEM). Details of the characterization can be found in our previous publications\cite{NJP}. Size of the nanocrystals used were obtained form TEM images as well as from the XRD results. For our experiment, we used samples with the average sizes varying between 15 nm to 100 nm for LCMO and between 35 nm to 130 nm for LSCO.
 
\noindent
We used annealed pellets of the nanocrystals for transport measurements. The grain sizes quoted here are the average size in the pelletized condition. Grain growth to larger sizes were carried out by controlling the annealing temperature. The low temperature magnetotransport measurements (0.3 K $<$ T $<$ 10 K) in a magnetic field upto 14 T were done using a  Physical Property Measurement System\cite{PPMS}. The high temperature resistivity measurements (T $>$ 10 K) were done using standard d. c. four probe technique in a closed cycle refrigerator. Though the paper focuses on the transport below 5 K, the measurements to higher temperatures were done for the sake of completeness. 

\noindent
In Fig. 1 (a) and (b) we show the magnetization data (M-H curves) at T = 2 K taken on both the LCMO and LSCO systems for the smallest sized nanocrystals.  The LCMO (15 nm) has a ferromagnetic $T_{C} = 260$ K and the LSCO (35 nm) has a $T_{C} = 198$ K. None of the samples show superparamagnetic behavior. However, they have a low temperature saturation magnetization that is smaller than that seen in the respective bulk samples.  The size dependence of the ferromagnetic transition temperature and magnetic behavior of both LCMO\cite{NJP} and LSCO\cite{TapatiJNN,Fuchs} systems have been studied in details before and are not discussed here.
In table I, we show the $T_C$ of the samples studied. The coercive field $\mu_0H_C$ at 2 K  for the LCMO sample is low and is typically $\leq 0.1$ T  while for LSCO, it is large and is $\approx$ 1 T. The magnetoresistance measurements have been done to fields much larger than $H_{C}$.

\begin{figure}[t]
\begin{center} 
\includegraphics[width=8cm,height=7cm]{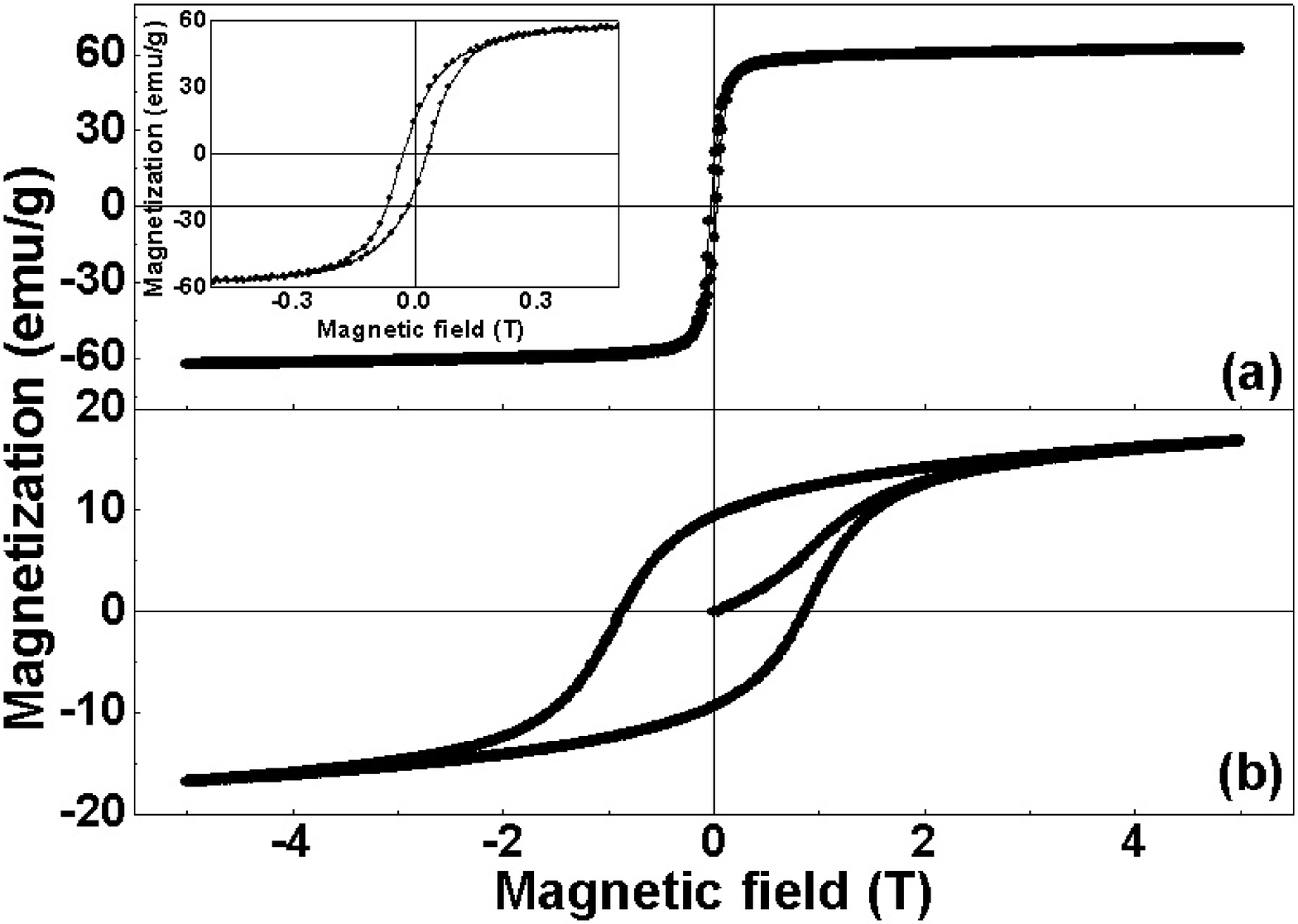}
\end{center}
\caption{M-H curves for (a) LCMO (d = 15 nm) and (b) LSCO (d = 35 nm) nanocrystals at 2 K. The inset in (a) shows the expanded region of the M-H curve of LCMO nanocrystals near the low field region to show the low value of the coercive field.} 
\label{Fig1}
\end{figure}

\begin{table}
\caption{\label{tab:table1}Average particle diameter (d) and ferromagnetic Curie temperature ($T_C$) of $\mathrm{La_{0.67}Ca_{0.33}MnO_3}$ samples.}
\begin{ruledtabular}	
\begin{tabular}{cc}
$d$ (nm)& $T_C$ from $M$-$T$ measurements (K)\\
\hline
15&242\\
23&260\\
50&294\\
102&269\\
\end{tabular}
\end{ruledtabular}
\end{table} 

\section{RESULTS}
The results of the transport and magnetotransport experiments are presented in different subsections. We first present the resistivity and magnetoresistivity  data below 10 K upto a field $\mu_{0}H = 14$ T. This is followed by the nonlinear conduction data (I-V curves) down to 0.3 K and in magnetic fields  upto 5 T. 
\subsection{ELECTRICAL TRANSPORT IN LCMO AND LSCO NANOCRYSTALS (T $<$ 5 K)}
In Fig. 2, we show the typical low temperature resistivity ($\rho$) versus temperature (T) curves for LCMO nanocrystals (d = 15 nm) taken in the presence of magnetic field. In the inset of Fig. 2, the variation of the magnetoresistance (MR) as a function of the applied magnetic field at two different temperatures, 0.3 K and 10 K, are shown. The MR is defined as $MR(\%) = (\frac{\rho (0)-\rho (H)}{\rho (0)})$ X 100. The data for other sizes are qualitatively similar and are not shown.

\begin{figure}[t]
\begin{center} 
\includegraphics[width=8cm,height=7cm]{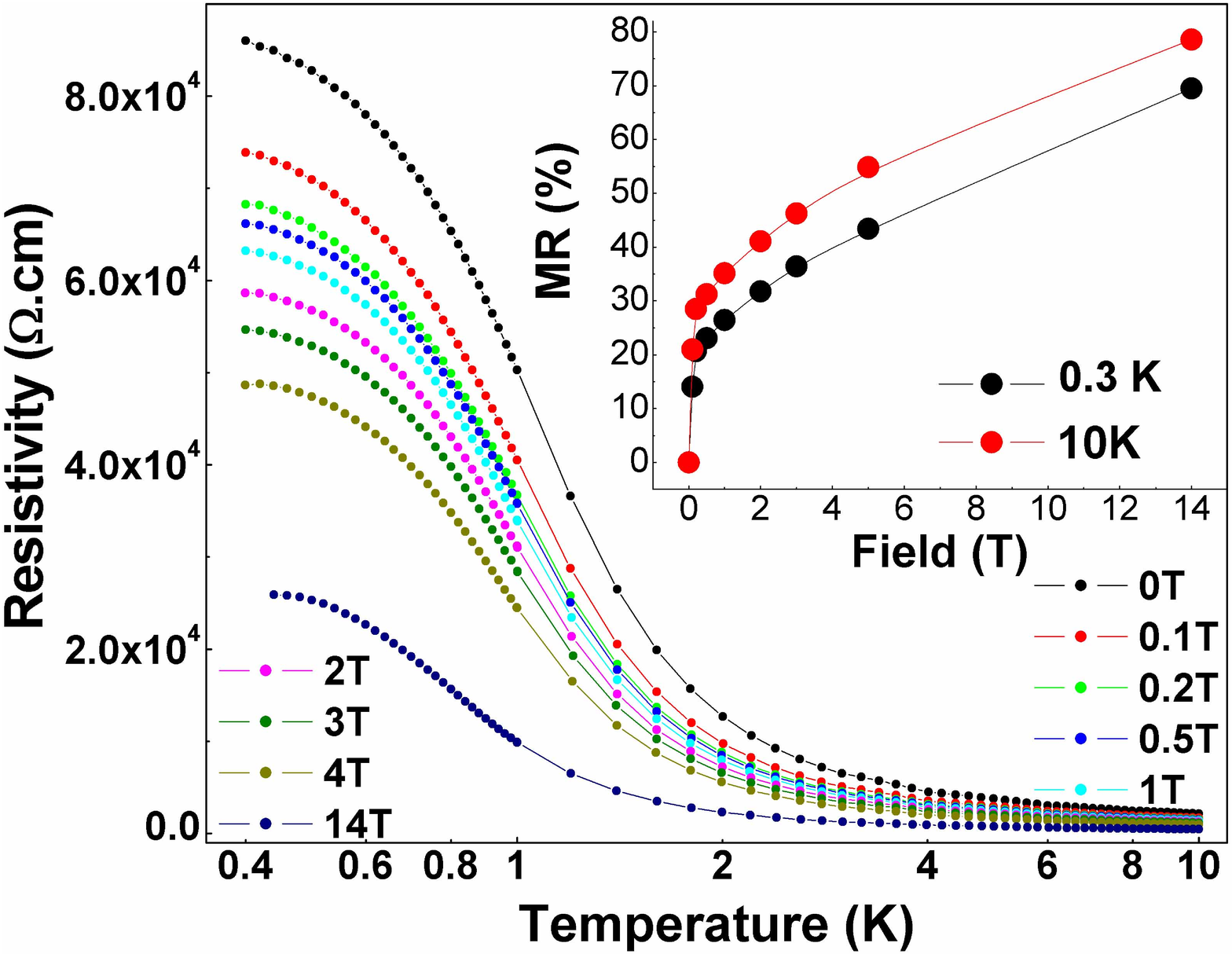}
\end{center}
\caption{Resistivity vs temperature plots for LCMO (d = 15 nm) nanocrystals taken under different magnetic fields. The inset shows the MR as a function of the applied magnetic field at T= 0.3 K and T = 10 K. (Color online).}
\label{Fig2}
\end{figure}

\noindent
The resistivity for the LCMO nanocrystals shows a pronounced upturn below 5 K. $\rho$ is strongly suppressed by the magnetic field and the magnitude of MR at low field ($\mu_0H < 0.2$ T) reaches a value of 25 $\%-30\%$. (Note: This kind of low field magnetoresistance (LFMR) effects have been observed in polycrystalline as well as nano-crystalline samples of LCMO before, albeit at higher temperatures\cite{Mahesh,Huang,Arunava}). At higher field, the MR increases in magnitude linearly without saturation till the highest field measured (14 T) and reaches a maximum magnitude of $\approx 70-80\%$. This behavior is more or less temperature independent below 10 K. 

\noindent
Typical resistivity as well as the MR data for  LSCO nanocrystals are shown in Fig. 3 for the sample with d = 35 nm. The resistivity of the LSCO system is much lower compared with that of the LCMO system, although it shows a clear rise with lowering of temperature. The MR in this material is lower. It shows no discernible LFMR and shows a linear increase (without saturation) from lowest field till the highest measurement field and reaches a magnitude of 25$\%$, which is also temperature independent at low temperatures. Absence of a strong LFMR in LSCO is due to the absence of a strong spin polarization in the LSCO system as discussed below and reflects an essential difference between the two systems.

\begin{figure}[t]
\begin{center} 
\includegraphics[width=8cm,height=7cm]{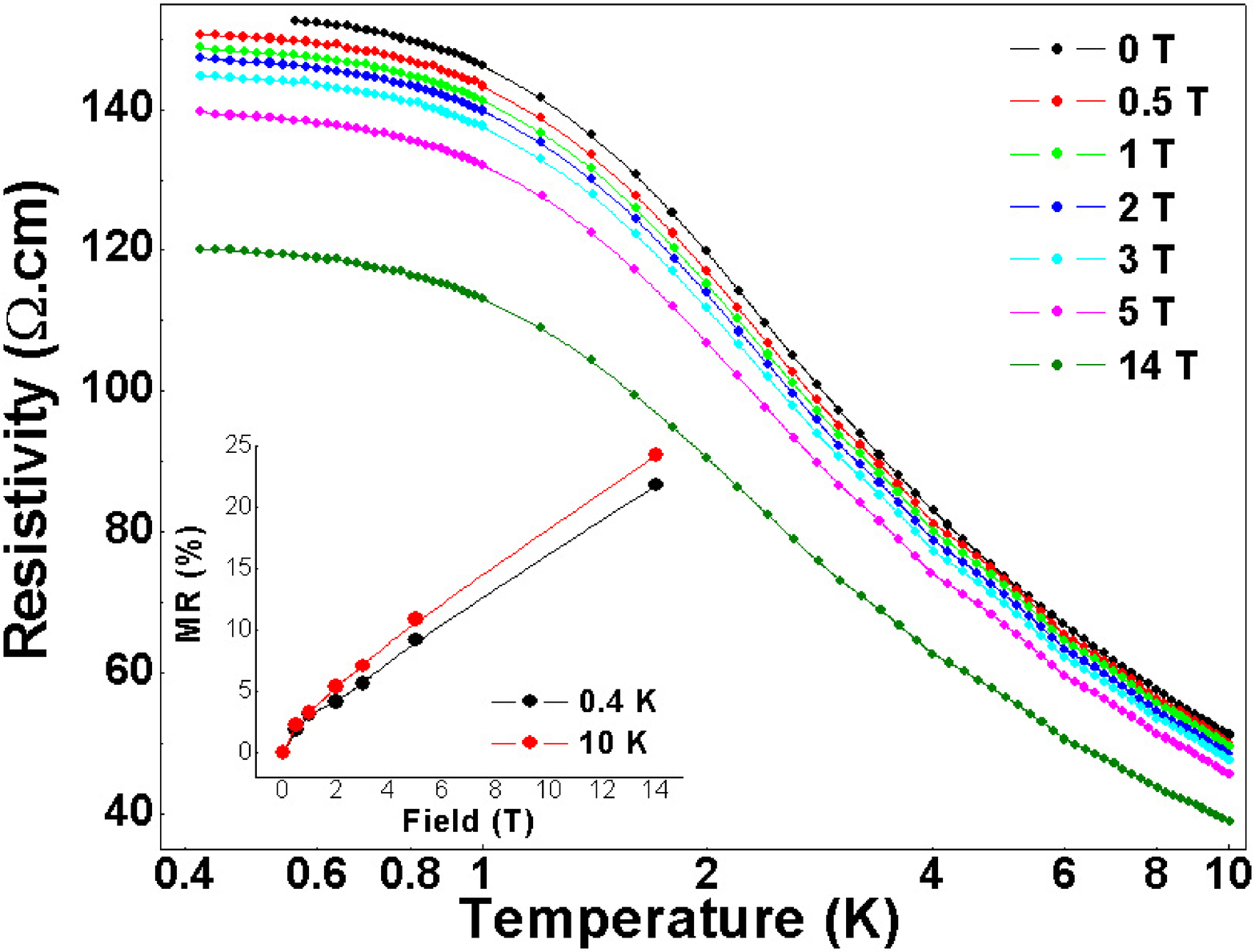}
\end{center}
\caption{Resistivity vs temperature plots for LSCO (d = 35 nm) nanoparticle taken under different magnetic fields. The inset shows the MR as a function of the applied magnetic field at T = 0.3 K and T = 10 K. (Color online).}
\label{Fig3}
\end{figure}

\noindent
In Fig. 4, we have plotted the observed resistivities of the two systems for the smallest sample sizes as ln$\rho$ vs. $T^{-1/2}$. As can be seen in Fig. 4, below 5 K there are two distinct regions. Down to T $\approx$ 1 K for LCMO and down to T $\approx$ 1.5 K for LSCO, the resistivity follows Eqn. 1. For T $<$ 1 K, there is a deviation form Eqn. 1 and the temperature dependence softens. This is the case for all the samples studied with different average sizes. The activation energy, $\Delta$, obtained from the above plot (evaluated in the temperature region where Eqn. 1 is valid) has a distinct dependence on the average size (d), which we discuss later on. The value of the activation energy is much less for the LSCO system compared to the LCMO system. For the smallest size sample in the LCMO system, the activation energy, $\Delta$, reaches a value of $\approx 2$ meV and the temperature range in which we determined the activation energy is much smaller than $\frac{\Delta}{k_{B}}$. For the LSCO system, the activation energy for the smallest sample is two orders of magnitude smaller and the value of $\frac{\Delta}{k_{B}}$ is smaller than the temperature range of measurement. Thus caution should be exercised in interpreting the resistivity in LSCO as an effective manifestation of CB.

\begin{figure}[t]
\begin{center} 
\includegraphics[width=8cm,height=7cm]{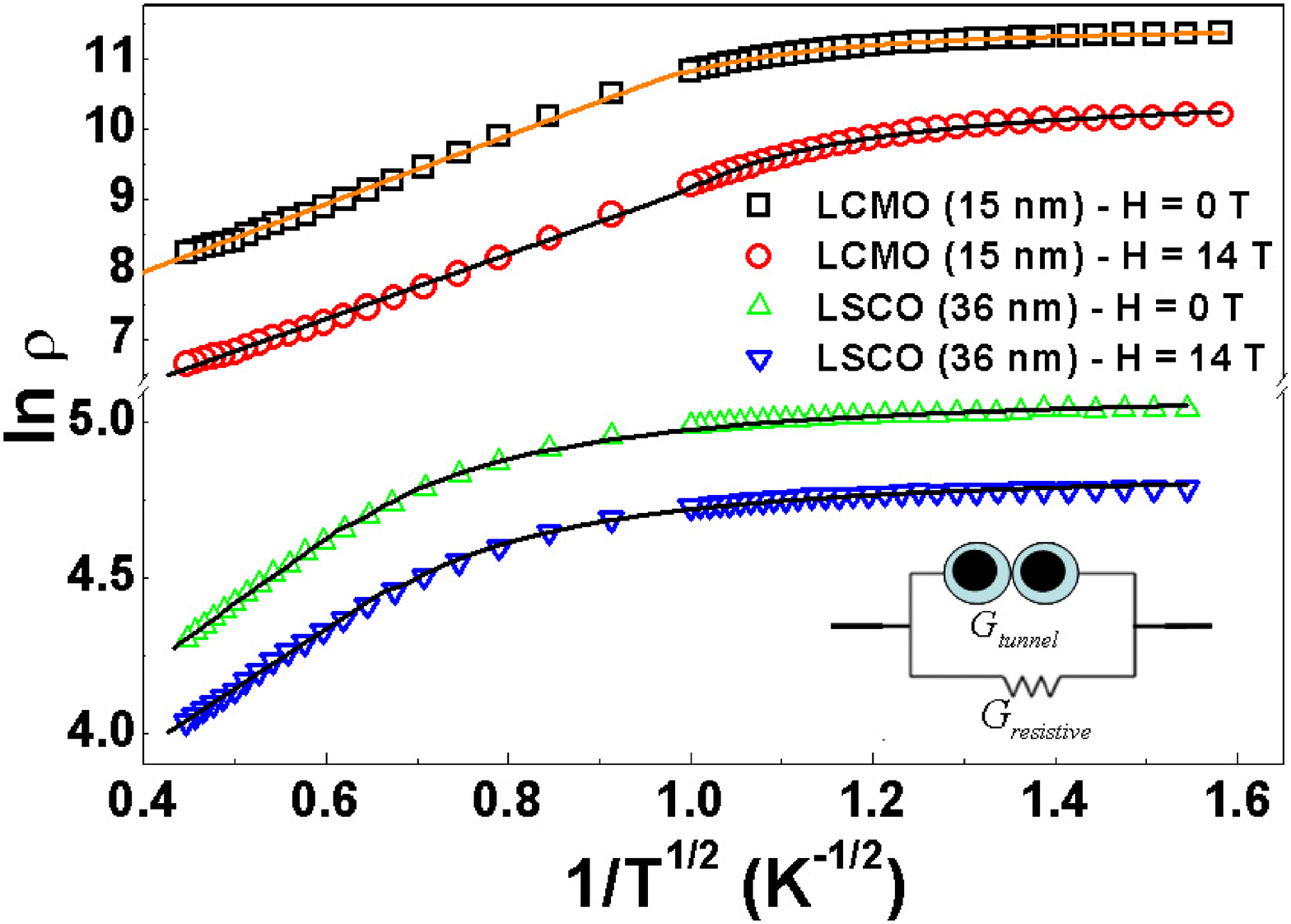}
\end{center}
\caption{ln$\rho$ vs $\frac{1}{T^{1/2}}$ for LCMO and LSCO nanocrystals for T $<$ 5 K. The solid lines are fits to the data according to Eqn. (2) (Color online).}
\label{Fig4}
\end{figure}

\noindent
For T $<$ 1 K, the temperature dependence of $\rho$ crosses over to a region with very soft T dependence as can be seen in Fig. 4. It appears that $\rho$ reaches a temperature independent value as T $\rightarrow$ 0. This is in contrast to what is expected for activated transport. To understand the temperature dependence, we propose a simple empirical model where the conduction takes place through a tunnelling as well as non-tunnelling path. The activated transport occurs through the mechanism of inter-grain tunnelling where the barrier is provided by the material in the grain boundary region. It is also likely that there are regions in the sample, where the grains may have better contact that can act as highly resistive but non-tunnelling paths. The schematic cartoon of the two channel model is shown as an inset in Fig. 4. The two paths have conductivity marked as $G_{tunnel}$ and $G_{resistive}$ respectively, and represent the lumped contributions from the two paths. At higher temperature, the current is mainly carried by the tunnelling paths with the resistive paths  carrying smaller current. As the temperature decreases, due to the activated nature of transport through the tunnelling paths, they carry less and less current and the non-tunnelling resistive path that has weaker temperature dependence, makes significant contributions. This leads to softening of the temperature dependence on cooling. We can quantify the simple model by the relation:
\begin{eqnarray}
\frac{1}{\rho} = \frac{A_i}{\rho_i} + \frac{A_m}{\rho_m}
\end{eqnarray}
where, $A_i$ and $A_m$ give the relative fractions of the tunnelling (insulating) phase and the metallic parts respectively in the sample. Here, the insulating tunnelling channels have been taken to have a resistivity which follows Eqn. 1, $\rho_{i} = \rho_{i0}exp (2\sqrt{\frac{\Delta}{k_BT}})$. The non-tunnelling resistive channels have the resistivity of the form of a disordered metal $\rho_m = \rho_{m0}-\beta T^n$ with weak temperature dependence. For a disordered metallic region with weak disorder, one would expect the exponent $n \approx 3/2$ when the scattering process is dominated by weak localization\cite{Lee}. The fits obtained using this model are also shown in Fig. 4 as solid lines.

\noindent
The low temperature resistivity of both the systems follow Eqn. 2 in the presence of magnetic field also (See Fig. 4). This has been investigated in details for the samples with the smallest particle sizes. The magnetic field, as we will see, affects both the parameters $\rho_{i0}$ and $\Delta$, thereby giving rise to the MR. The field  dependence of  $\Delta$ and $\rho_{i0}$ (obtained by fitting the data to Eqn. 2) are shown in Fig. 5 and Fig. 6  for LCMO and LCSO respectively.

\begin{figure}[t]
\begin{center} 
\includegraphics[width=8cm,height=7cm]{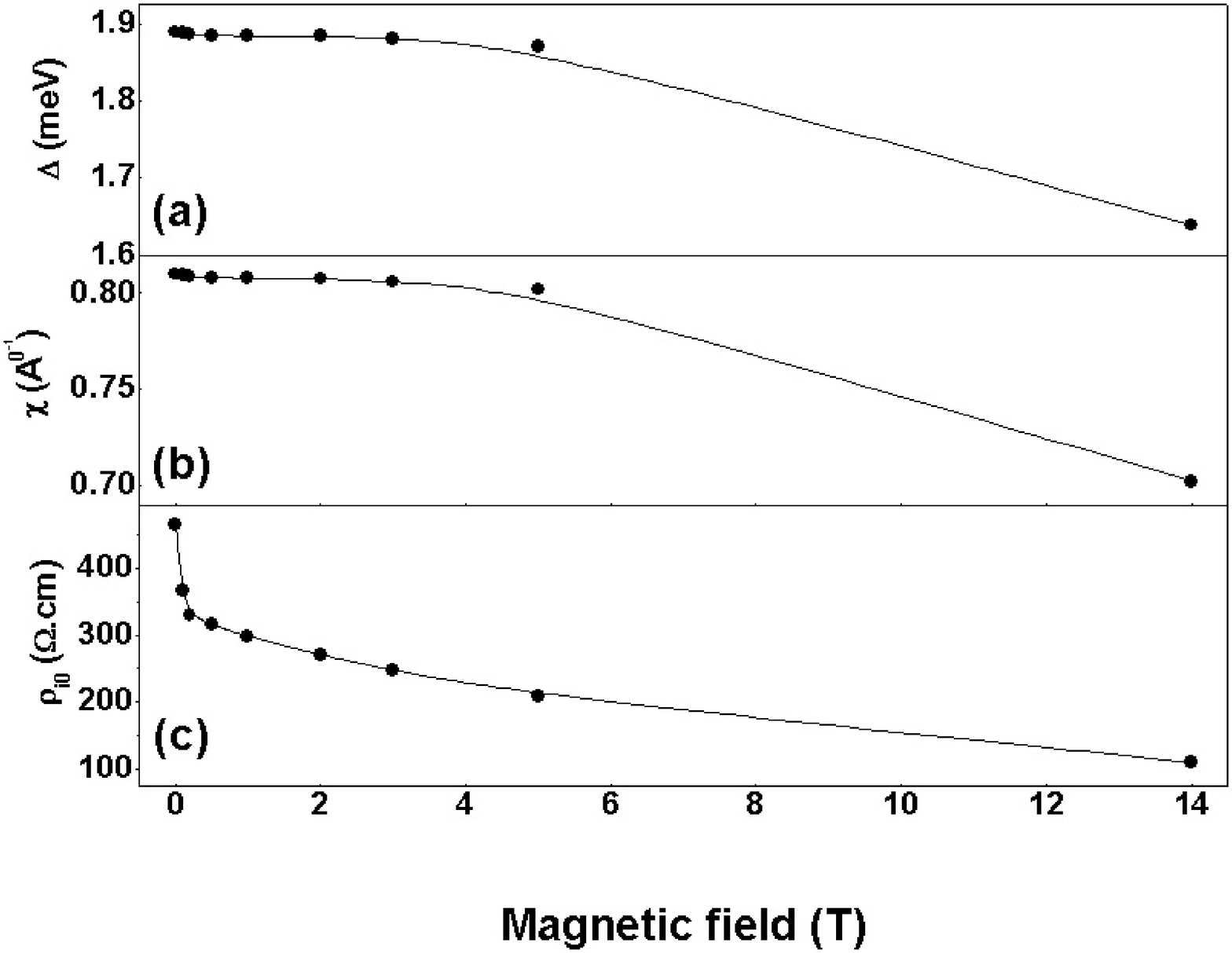}
\end{center}
\caption{Variation of (a)$\Delta$, (b)$\chi$ and (c)$\rho_{i0}$ with applied magnetic field for LCMO (d = 15 nm) nanocrystals.}
\label{Fig5}
\end{figure}

\begin{figure}[t]
\begin{center} 
\includegraphics[width=8cm,height=7cm]{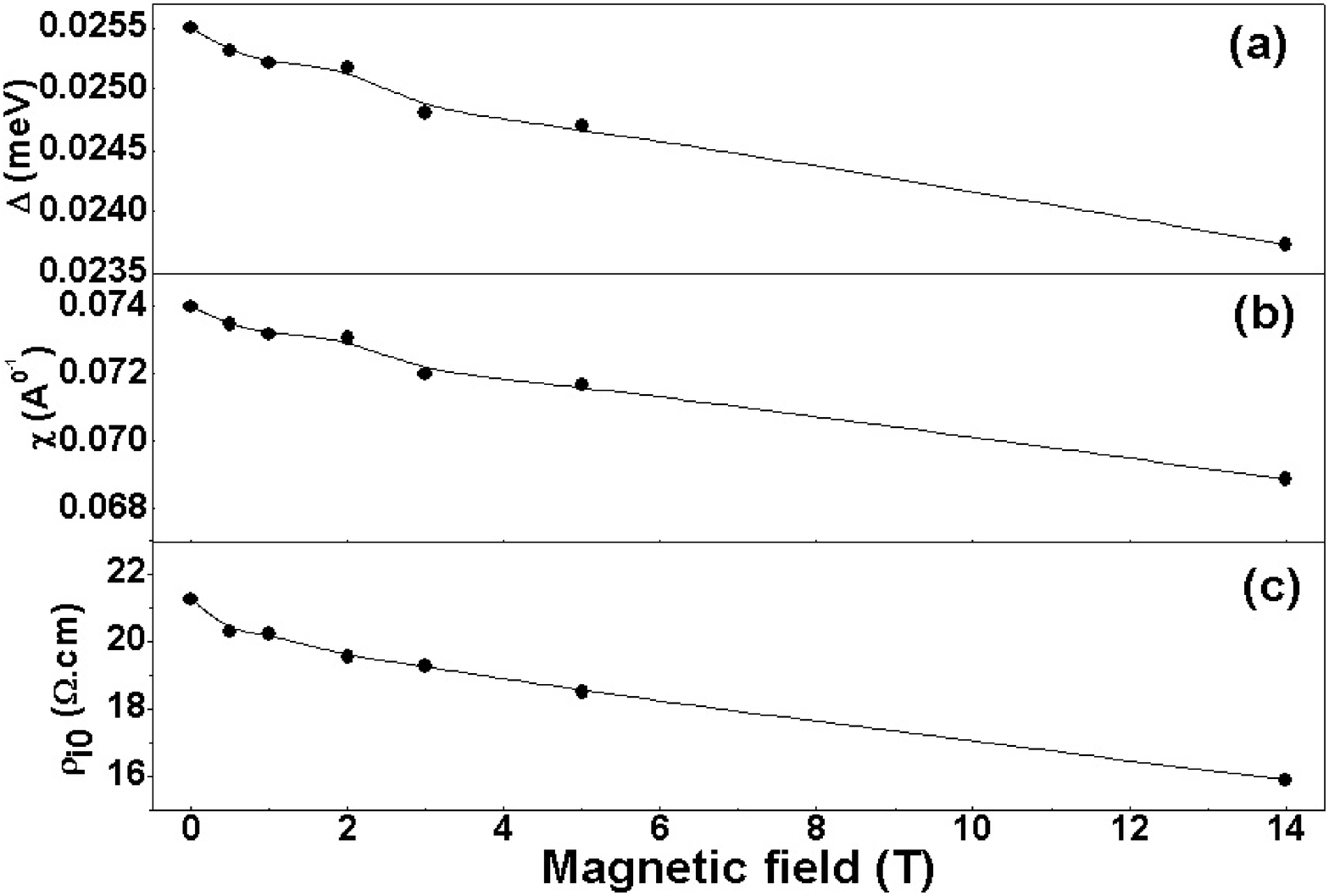}
\end{center}
\caption{Variation of (a)$\Delta$, (b)$\chi$ and (c) $\rho_{i0}$ with applied magnetic field for LSCO (d = 35 nm) nanocrystals.}
\label{Fig6}
\end{figure}

From Fig. 5 and Fig. 6, the essential differences of the two systems can be seen. For the LCMO sample, which shows the two stages of MR, it can be seen that the low field and the high field regions have clearly distinct features, and are thus, expected to have different origin. The low field rapid change in MR occuring for H $<$ 0.2 T (LFMR) arises from a change in the prefactor $\rho_{i0}$. In this range, the activation energy $\Delta$ remains more or less field independent. $\Delta$ changes only for H $>$ 4 T.  These changes in $\Delta$ as well as the gradual change in $\rho_{i0}$ do not saturate even at a field of 14 T. In LSCO that has no LFMR, there is a gradual change in both the parameters $\Delta$ as well as $\rho_{i0}$. In the temperature range where the transport is dominated by tunnelling in Coulomb blockade regime, the identification of the origin of the two regions in MR through field dependences of the two parameters, namely $\rho_{i0}$ and $\Delta$, have not been reported before. 

\noindent
\subsection{NON-LINEAR TRANSPORT (I-V CHARACTERISTICS) OF LCMO and LSCO NANOCRYSTALS BELOW 5 K}
The I-V curves for the smallest size LCMO sample (d $\sim$ 15 nm)  are shown in Fig. 7 for (a) different temperatures at $\mu_{0}H$ = 0  and (b) at T = 0.3 K for different magnetic fields upto 5 T. The I-V curves for the LCMO nanocrystals show very strong non-linearity. It can be seen that magnetic field has a strong effect on the I-V curves particularly at the lowest temperature, where the magnetic field changes the nature of the I-V curves significantly. The strong MR seen is reflected in the field dependence of the I-V curves. At higher temperatures and in particular above 1 K, the non-linearity becomes weaker and the I-V curves become more linear.  In contrast, the application of the magnetic field enhances the non-linearity significantly. For instance, at a bias of 50 mV, the current in a field of 5 T is more than an order higher than that at zero field. 

\begin{figure}[t]
\begin{center}
\includegraphics[width=8cm,height=10.5cm]{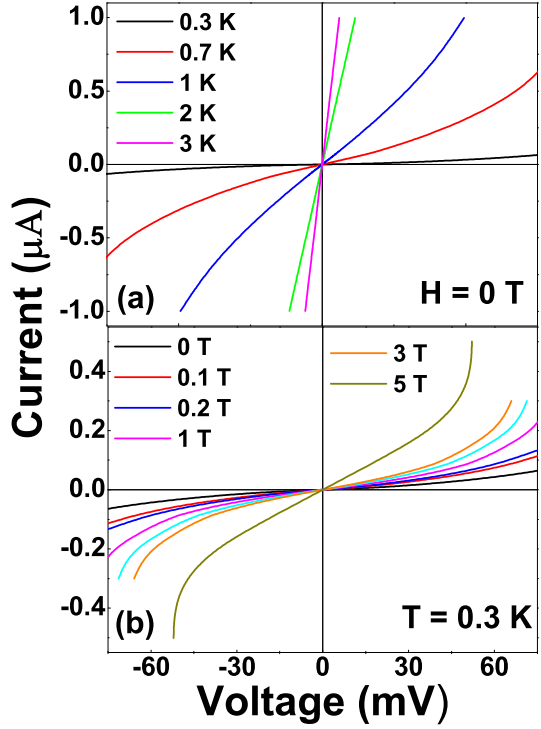}
\end{center}
\caption{I-V curves of LCMO nanocrystals (d $\sim$ 15 nm) at (a) $\mu_{0}H$ = 0 T for different temperatures and (b) at T = 0.3 K for different magnetic fields (Color online.)}
\label{Fig7}
\end{figure}

\noindent
Similar data for the LSCO nanocrystals are shown in Fig. 8(a) and 8 (b). The I-V curves for the two systems show qualitatively different behavior. For the LSCO system, the I-V curves show a small non-linearity for bias $<$ 20 mV upto a temperature of 2 K. At higher bias and for higher temperatures, the curves are linear. The I-V data does not seem to have any significant magnetic field dependence. The low value of the resistivity of the LSCO nanocrystals as well as the low value of $\Delta$ show that the tunnelling is not well developed in the system in the temperature range we are working. Presumably at much lower temperatures, when T$< \frac{\Delta}{k_{B}}$, we may expect to see the strong non-linearity seen in the LCMO system. 

\begin{figure}[t]
\begin{center}
\includegraphics[width=8cm,height=7cm]{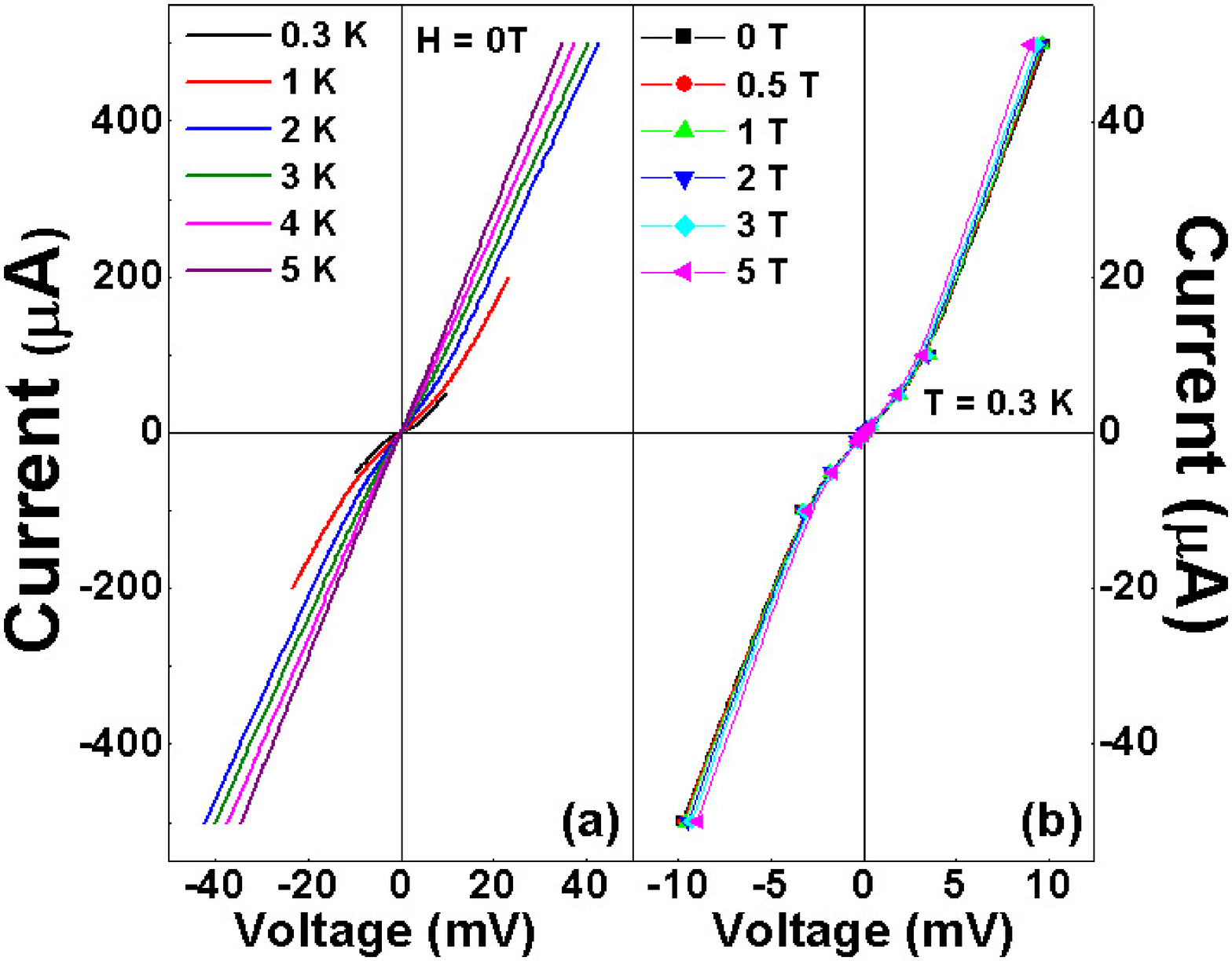}
\end{center}
\caption{I-V curves of LSCO nanocrystals (d = 35 nm) taken at (a) $\mu_{0}H$ = 0 T for different temperatures and (b) at T = 0.3 K for different magnetic fields (Color online).}
\label{Fig8}
\end{figure}

\noindent
\subsection{ELECTRICAL TRANSPORT AT HIGHER TEMPERATURES (T $>$ 5 K)}
The main focus of this paper is the temperature range below 5 K. We report the data for T $>$ 5 K briefly here for the sake of completeness. The electrical resistivity in LCMO nanocrystals for different sizes for T $>$ 5 K are shown in Fig. 9. The transport in manganite nanocrystals above 10 K has been investigated extensively before\cite{garcia,Yuan,JPCM,SSC}. The data for LCMO shows that as the particle size is decreased, the absolute value of $\rho$ increases and for size below 50 nm, the resistivity has a negative temperature coefficient over the whole temperature range. The resistivity for the samples with smaller crystallite sizes show the predominance of the grain boundary in the transport process. We have reported before an extensive study of electrical transport in LSCO nanocrystals in the temperature range 2 K $<$ T $<$ 300 K\cite{TapatiJNN}.  We do not repeat the data for LSCO nanocrystals here. In the case of LSCO also, for nanocrystals with size below 60 nm, the resistivity shows negative temperature coeffcient over the whole temperature range.

\begin{figure}[t]
\begin{center}
\includegraphics[width=8cm,height=7cm]{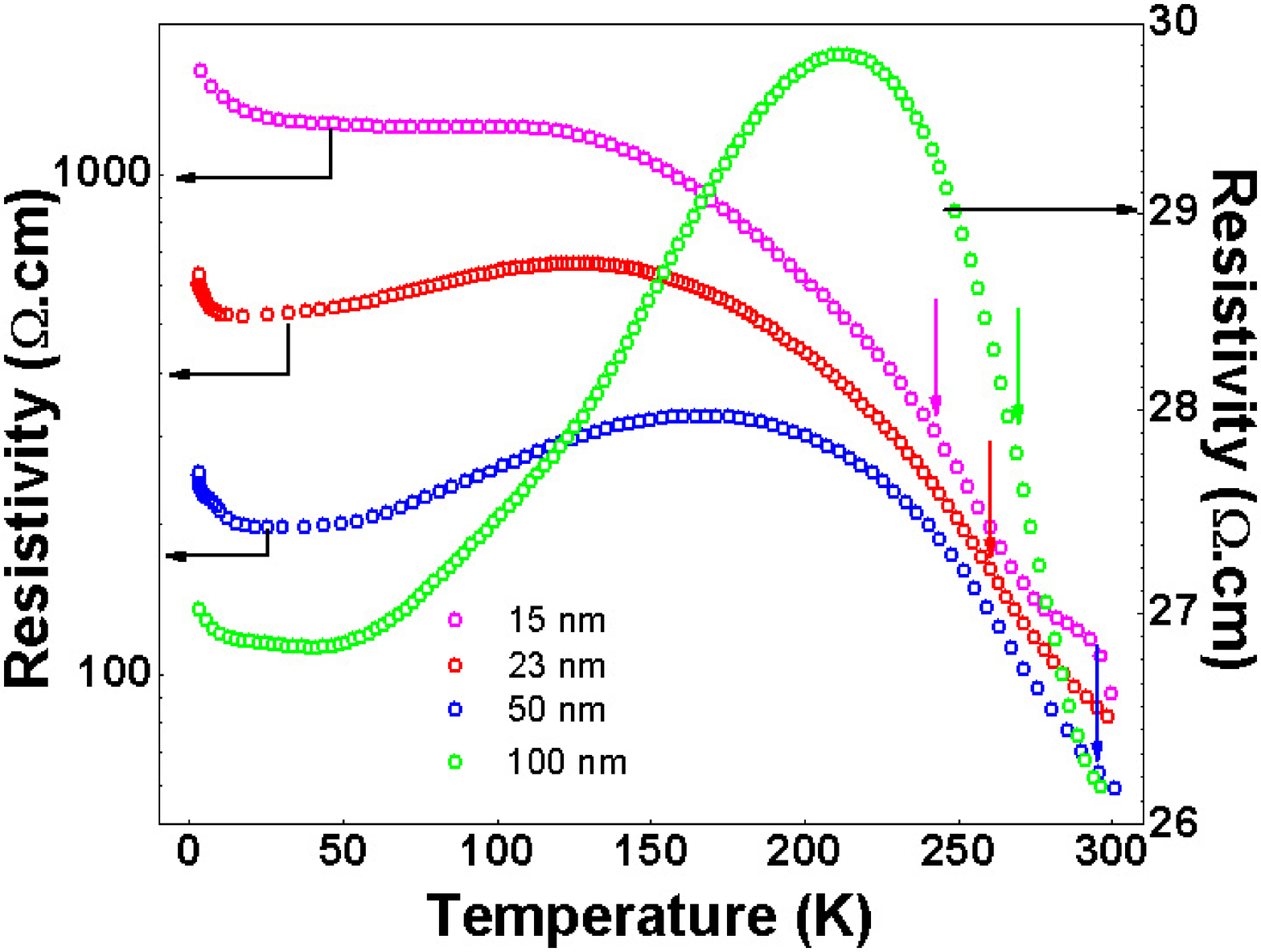}
\end{center}
\caption{High temperature resistivity as a function of T for LCMO nanocrystals of different sizes. The arrows mark the ferromagnetic transition temperature $T_{C}$ (Color online).}
\label{Fig9}
\end{figure}

\section{DISCUSSION}
In this section, we discuss the physical significance of the main observations and would try to obtain relevant information from the parameters as obtained from the data. The analysis of the data in the Coulomb blockade dominated low temperature regime below 5 K, has been done mainly using theories developed for transport in granular metals and also theories of spin polarized tunnelling. One of the important observations that we have made is the effect of magnetic field on the non-linear conductance which specifically occurs at the lowest temperatures, T $<$ 1 K, and also the observation of a gap like feature in the bias dependent conductance as shown below.  

\subsection{TRANSPORT IN THE COULOMB BLOCKADE REGIME AND SIZE DEPENDENCE OF $\Delta$} 
The measurements done to low temperatures allow us to obtain the value of the activation energy $\Delta$ unambiguously. $\Delta$ for transport in a granular medium at low temperatures is related to the Coulomb charging energy $E_C$ of a metallic grain, although they are distinctly different quantities\cite{Abeles1}. In our case, the metallic grains (characterized by an average size $\mathrm{d}$) are the ferromagnetic nanocrystals of LCMO and LSCO. The Coulomb charging energy $E_{C}$ $\propto \frac{1}{d}$. At low temperatures the transport from one grain to the other is by tunnelling through the insulating grain boundary regions, which has an average thickness $\mathrm{s}$, which is the average grain separation. The activation energy $\Delta$ is a function of $\frac{s}{d}$ and is  given by\cite{Abeles1} :
\begin{eqnarray}
\Delta = 8\chi e^2\frac{(s/d)^2}{1+2(s/d)} 
\end{eqnarray}
Here, $\chi$ is the inverse decay length of the tunnelling wave function in the barrier which is related to the barrier height at the surface of the grain. We plot $\Delta$ in Fig. 10 as a function of the inverse particle size ($1/d$) and fit the experimental points (for the LCMO nanocrystals) using Eqn. 3. From the fit, we can extract two important parameters $\frac{s}{d}$ and $\chi$. Since average $\mathrm{d}$ is known, average $\mathrm{s}$ can be obtained form this fit. From the fit, we obtain for LCMO, $s = 4.6 \pm 0.3$ nm, and $\chi = 0.81 \pm 0.06 {\AA}^{-1}$ for the data at zero field. We would like to explore whether we can physically justify the size of $\mathrm{s}$. The main assumption of the model used here is that the tunnelling barrier is provided by the inter-grain region.  This, for a compacted system of nanocrystals, will be the average inter-grain separation which in turn will be determined mainly by the width of the grain boundary. In nanocrystals like LCMO, it has been established that there is a disordered shell of thickness $\delta$ at the surface of the nanocrystals\cite{NJP}. The particle separation ($s$) is thus expected to be $\approx 2\delta$.  The value of $\delta$ in these samples has been estimated independently from the size dependence of the magnetization measurements\cite{NJP}, which gives $\delta \approx 2.47$ nm so that $s \approx$ 4.9 nm, which is very close to what we have obtained from the fits to Eqn. 3. (The samples used in reference 26 and this investigation are the same). Recent TEM studies\cite{Curiale} have also shown the existence of a disordered shell having a thickness of approximately 2 nm in manganite nanoparticles. It is thus encouraging that there is indeed quantitative confirmation of the important parameter $s$ as obtained from very different experiments. Eqn. 3 is a two-parameter equation. The fit to eqn. 3 was done using a non-linear fit method that gives best fit to the data and the parameters need be physically meaningful. 

\begin{figure}[t]
\begin{center}
\includegraphics[width=8cm,height=7cm]{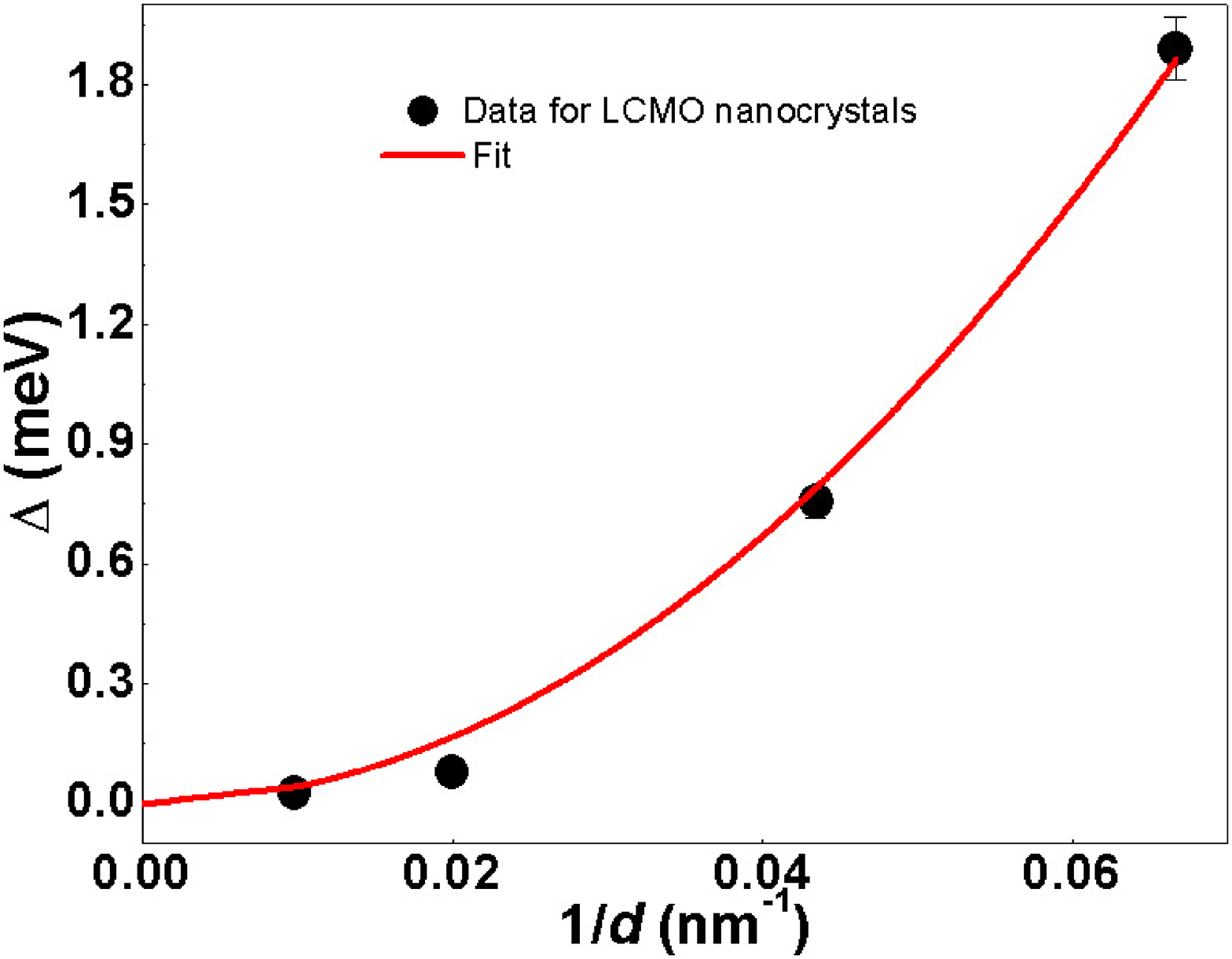}
\end{center}
\caption{$\Delta$ vs $\frac{1}{d}$ for LCMO nanocrystals. The solid line is a fit to the equation for $\Delta$ (Eqn. 3) (Color online).}
\label{Fig10}
\end{figure}

\noindent 
The inverse decay length $\chi$ is related to the potential barrier to tunnelling by the relation, $\chi =[\frac{2m^{*}\Phi_{B}}{\hbar^{2}}]^{1/2}$ where barrier height $\Phi_{B}= V_{eff}-\mu$, $V_{eff}$ being the barrier potential, $\mu$ is the chemical potential and $m^{*}$ is the effective carrier mass. Using the value of the electron mass as the carrier mass, we could get an estimate of the potential barrier $\Phi_{B} \approx 25$ meV.  

\noindent 
The value of $\Delta$ obtained for the LSCO system is very low and it may not be advisable to analyse its size dependence. Nevertheless, we can use Eqn. 3 (albeit with caution) for the data on the smallest nanocrystals to obtain the parameters $\mathrm{s}$ and $\chi$. We obtain, $s = 9.4 \pm 0.2$ nm and $\chi = 0.074 \pm 0.002 {\AA}$. The thickness of the non-magnetic shell $\delta$, can be estimated from the size dependence of the saturation magnetization of the nanocrystals\cite{TapatiJNN}. We estimate $\delta \approx$ 4.3 nm for the smallest nanocrystals of LSCO. This gives an independent estimate of $s \approx 2\delta \approx 8.6$ nm, which is very close to $s = 9.4 \pm$ 0.2 nm as obtained from the transport data. 

\noindent
Analysis of the size dependence of the activation energy $\Delta$ is very useful because it gives values for important parameters that govern the tunnelling process. This analysis also vindicates the basic assumption that in these systems, the non-magnetic and insulating shell on the nanocrystal  serves as the tunnel barrier. This is an important consideration because in this system, unlike transition metal/insulating oxide systems, no extra insulating phase has been added.  
 
\noindent
The charging energy $E_C$, though related to the activation energy $\Delta$, is a distinct physical quantity. Its value depends not only on the particle size $\mathrm{d}$, but also on the interparticle separation $\mathrm{s}$. This is given as\cite{Abeles1}: 
\begin{eqnarray}
E_C = (\frac{e^2}{\pi\epsilon_0\epsilon_d})(\frac{s}{s+d/2})    
\end{eqnarray}
where $\epsilon$ is the dielectric constant of the medium, which for the manganites, is taken as $\approx$ 10 (Ref. 12). Using the known value of $\mathrm{s}$ from the previous analysis, we obtain $E_C\approx$ 7 meV for the smallest LCMO particle and 0.25 meV for the largest LCMO particles ($d \approx$ 100 nm). Thus, even for the largest particles, the lowest temperature range of measurement is lower than $E_C$. The basic assumption of the validity of transport in CB regime is thus satisfied.

\subsection {MAGNETORESISTANCE AND DEPENDENCE OF TRANSPORT PARAMETERS ON MAGNETIC FIELD}
The field dependence of the transport parameters, $\Delta$, $\rho_{i0}$ and $\chi$ for the LCMO nanocrystal system with smallest size ($d \approx$ 15 nm) are shown in Fig. 5. The changes in these parameters cause the suppression of the resistance in the magnetic field and hence the MR. The field dependences of $\Delta$ and the resistivity pre-factor $\rho_{i0}$ are distinct, and, as stated before, distinguish the physical origin of the LFMR (as mainly arising from a reduction in $\rho_{i0}$) and the MR at higher field (as mainly arising from a reduction in $\Delta$ as well as a gradual reduction in $\rho_{i0}$). In the same graph, we also show the field dependence of $\chi$ obtained from Eqn. 3. The important observation in LCMO is the existence of a rather substantial LFMR when the transport has predominantly tunnelling nature. This is traced to a sharp lowering of $\rho_{i0}$ in low magnetic field.  The term $\rho_{i0}$ depends on the spin polarization of the tunnelling electrons and relative orientation of the magnetization of the two nanocrystals between which the tunnelling occurs. It has been shown that the tunnelling conductance for spin-polarized tunnelling $\propto (1+P^{2}cos\theta)e^{-2\chi s}$, where $\mathrm{P}$ is the spin polarization and $\theta$  is the angle between the  moments  of two ferromagnetic nanocrystals~\cite{Huang}. In that case the  value of $\rho_{i0}$ due to spin polarization will be: 
\begin{eqnarray}
\rho_{i0} \propto P^{2}(P^{-2}-cos \theta_{avg})e^{2\chi s}
\end{eqnarray}
where, $cos\theta_{avg}$ is the average of the $cos\theta$ factor. Due to the high value of the spin polarization of the tunnelling electron and the alignment of the magnetic moments in the nanocrystals  at low field, $cos\theta_{avg}$ will increase substantially when the field is applied, leading to a decrease in $\rho_{i0}$ as observed in the low field region. This decrease in $\rho_{i0}$ due to spin polarization is expected to saturate at field values where the technical saturation point is achieved in the M-H curves. At this field value, the nanocrystals will have their spins almost aligned. The continual and gradual decrease of $\rho_{i0}$ without saturation as the field is enhanced to 14 T  arises from decrease of $\chi$  as discussed below.

\noindent
From Eqn. 3 it can be seen that since $\frac{s}{d}$ does not change with magnetic field, the change in the activation energy $\Delta$ will arise from a change in the parameter $\chi$ which is a measure of the barrier height $\Phi_{B}$. The reduction in $\chi$ (and hence $\Delta$) in magnetic field is thus expected to arise from a reduction in $\Phi_B$. We find that the changes in $\chi$ in magnetic field do not saturate at higher field and thus provide a clear basis for nonsaturation of the MR. The change in $\Phi_B$ in applied field, as derived from $\chi$, is shown in Fig. 11. At high field (fields above few T), the spins in the core region of the nanocrystals are aligned and the magnetization is mainly saturated. Any change in the barrier can arise from a change in the Zeeman energy. For LCMO, using the core spin value as the value of the saturation magnetization of the bulk $M_S \approx 3.6\mu_B$\cite{NJP}, the Zeeman energy at a field of 14 T is $\sim$ 6 meV. This is very close to the observed change in the barrier $\Phi_B$ (see Fig. 11). Thus the change in $\Delta$ and its non-saturation in high magnetic field can be traced to lowering of barrier $\Phi_B$ in magnetic field due to Zeeman effect and its consequent effect of suppression of $\chi$. The gradual lowering of the resistivity factor $\rho_{i0}$, which is $\propto e^{2\chi s}$ (see Eqn. 5) can also be explained by the lowering of $\Phi_B$ in magnetic field. Reduction in $\Phi_B$ lowers $\chi$ and leads to the depression of $\rho_{i0}$.

\begin{figure}[t]
\begin{center}
\includegraphics[width=8cm,height=7cm]{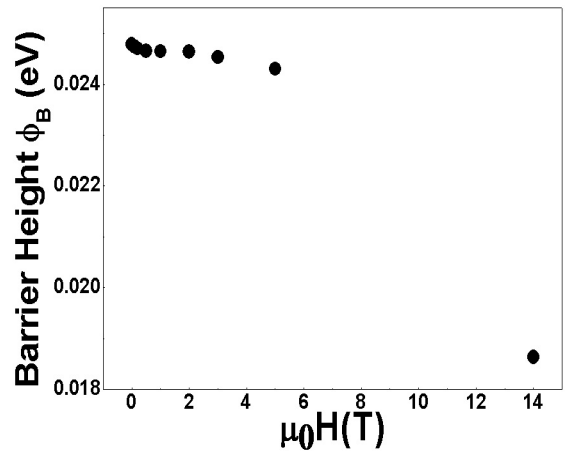}
\end{center}	
\caption{Dependence of barrier height $\Phi_{B}$ on magnetic field.}
\label{Fig11}
\end{figure}

\noindent
A quantitative evaluation of all the parameters, as is done in this investigation, allows us to cleanly separate out the effects of spin-polarized tunnelling and the effect of Coulomb blockade dominated tunnelling parameters in determining the low temperature transport in these materials. In particular, distinct effects of the magnetic field in the low field region (controlling the spin polarization dominated regime) and in the high field region (as arising from lowering of barrier) can be clearly evaluated. Extending these measurements to low temperatures and to high fields adds to this understanding.
\subsection{NONLINEAR I-V CHARACTERISTICS}
The analysis of the resistivity as well as the magnetotransport data has shown that in the temperature range of our investigation, the transport is through a network of tunnel junctions that consists of the metallic ferromagnetic oxide nanocrystals as electrodes and the intervening grain boundary regions as tunnel barriers. Thus, an investigation of the non-linear I-V characteristics in these materials at low temperatures will provide information on the tunnelling type transport through these materials. 

\noindent
The I-V characteristics of the systems studied have been shown in Fig. 7 and Fig. 8. The data on LSCO shows mainly ohmic behavior and the I-V curves are more or less linear. We do not discuss this system further. We plot the differential conductance $g = \frac{dI}{dV}$ vs bias voltage V curves in Fig. 12 for the LCMO system for the smallest nanocrystal size (d = 15 nm). The data shown are for zero field and with varying temperature and for fixed temperature (T = 0.3 K) with applied fields varying upto 5 T. 

\begin{figure}[t]
\begin{center}
\includegraphics[width=8cm,height=10.5cm]{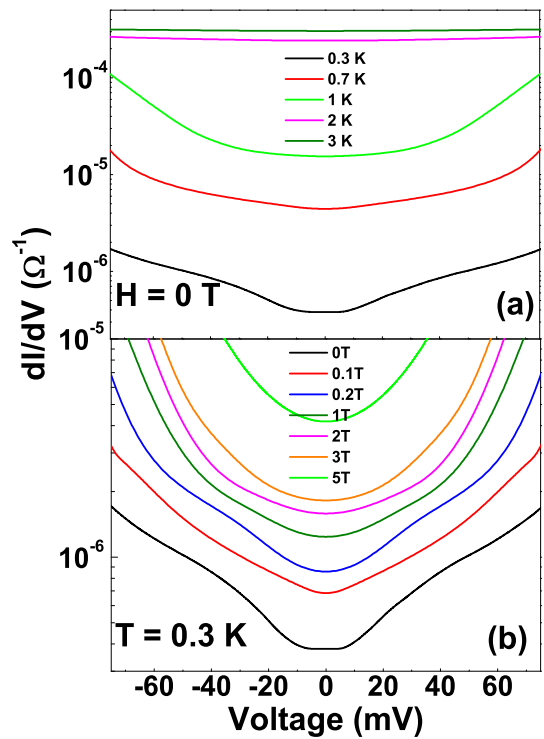}
\end{center}
\caption{The differential conductance vs V curves obtained from the curves shown in (a) Fig. 7(a) and (b) Fig. 7(b). (Color online).}
\label{Fig12}
\end{figure}

\noindent
It is noted that the $g$ at lowest temperatures show gap like features below a bias of 5 mV. The appearance of a gap of this magnitude in the $g-V$ curve is expected in view of the existence of a Coulomb charging energy $E_C$ which, for the nanocrystals of this size, we estimated to be $\sim$ 7 meV. The observed value and the estimates are thus very close.  Thus the non-linear transport data clearly shows existence of the Coulomb gap. Application of magnetic field as well as increase in temperature fills up the gap region and it closes by a field of 0.2 T and for T $>$ 0.7 K. The observation of such a clear gap like feature in the $g-V$ curve has not been seen before. This is because the present investigation was carried down to a temperature range, where the condition T $\ll \frac{E_C}{k_B}$ is well satisfied.

\noindent
To explore the nature of non$-$linearity in the I-V characteristics, the bias dependence of the differential conductance ($g$ versus V curves) have been fit to the empirical equation as has been done for investigating artificial grain boundary junctions before\cite{Mandar}:
\begin{eqnarray}
g(V,T,H) = g_0(T,H)+\Delta g(V,T,H) \\= g_0(T,H)+g_{\alpha}(T,H)\mid V\mid ^{\alpha}.
\end{eqnarray}
This is a two parameter relation that quantifies the nonlinearity through the exponent $\alpha$ and $g_{\alpha}$, the latter being a measure of the weight of the nonlinear transport. At lowest $T$, the fit was done for bias $>$ the gap value. $g_{\alpha}$, $\alpha$ and $g_0$ are all functions of T and H. The values of the parameters obtained from the $g(V)$ vs V curves are shown in Fig. 13 as a function of T and H for the LCMO nanocrystals of size 15 nm. The data have been fit to a bias ($\mathrm{V}$) upto $\sim$ 40 meV. For applied bias beyond this value, the data deviates strongly and shows considerable high conductance. We discuss this observation separately later on.

\begin{figure}[t]
\begin{center}
\includegraphics[width=8cm,height=7cm]{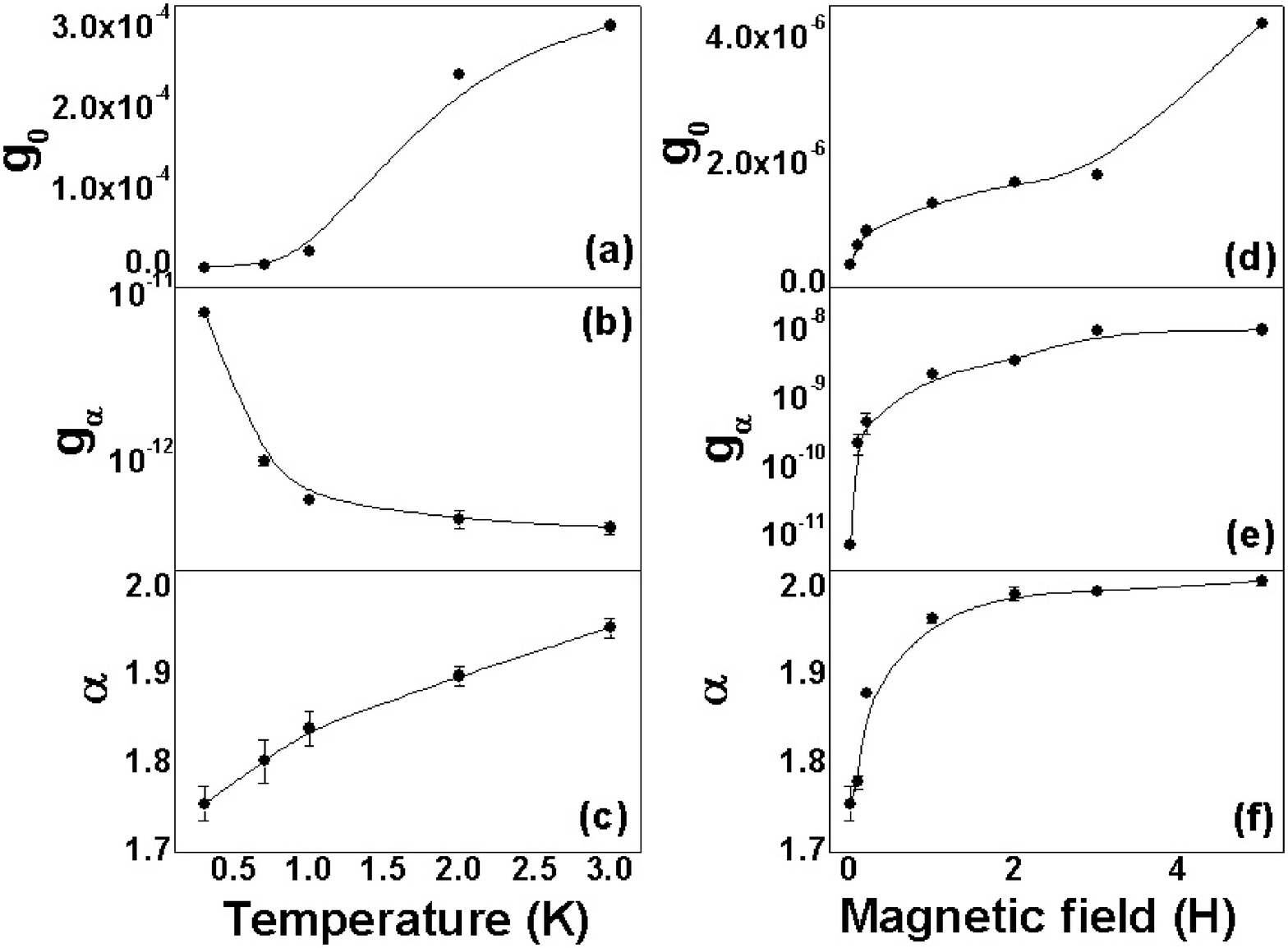}
\end{center}
\caption{Temperature dependence of the fit parameters (a) $g_0$, (b) $g_{\alpha}$ and (c) $\alpha$ in zero field, and magnetic field dependence of the fit parameters (d) $g_0$, (e) $g_{\alpha}$ and (f) $\alpha$ at T = 0.3 K for LCMO (d = 15 nm) nanocrystals.}
\label{Fig13}
\end{figure}

\noindent
The temperature and field dependent exponent $\alpha$ indicates the nature of the dominant tunnelling process. In principle, the transport through such grain boundary based junctions can have contributions from a number of processes. For conventional tunnelling through a barrier in a metal-insulator-metal (MIM) junction, when the electron does not suffer any scattering in the barrier, $\alpha$ = 2\cite{simon,wolf}. $\alpha$ decreases as the electron crosses the barrier through the localized states in the barrier. Specifically, $\alpha = \frac{n+2}{n+1}$, where, $n$ is the number of localized states in the barrier through which tunnelling occurs. For no scattering in the barrier, $\mathrm{n}$ = 0 and $\alpha$ = 2. For thick barriers often $n$ can be very large and in that case $\alpha \rightarrow$ 1. Past investigations on artificial GB junctions have shown the applicability of the above ideas for a GB based junction\cite{Hofener,Mandar}. It was also shown that at low temperatures, existence of disordered materials in the barrier region can contribute a $V^{1/2}$ term to $g(V)$\cite{Mandar}. Existence of all the parallel channels can contribute simultaneously to the tunnelling conductance $g(V)$ and the observed $\alpha$ will be an effective average value that will weigh the contributions from all the bias dependent channels. As the relative contributions of all the channels may vary with T as well as H, the exponent $\alpha$ can also vary. The observed value of the exponent $\alpha$ changes from a value of 1.7 at lowest temperature measured (and in zero field) and reaches a value of nearly 2 at higher temperature (as well as in higher applied magnetic field) as expected for a conventional barrier tunnelling junction\cite{simon}. We suggest that at lower temperatures, there is a contribution like $g(V) \propto V^{1/2}$ arising from a parallel transport through the disordered region (non-tunnelling type) which will lower the effective $\alpha$. The gradual change of the exponent $\alpha$ to the value 2 thus shows the predominance of the tunnel type transport through the barrier in the GB region. 
 
\noindent
The temperature dependence of the zero bias conductivity $g_0$ is expected to be similar to that of the conductivity (=$\frac{1}{\rho}$). $g_0$ increases as the temperature is increased from 0.3 K to 3 K and we find that down to 0.7 K, it follows the temperature dependence $g_0 \propto e^{-2\sqrt{\Delta_g/k_BT}}$. The value of the activation energy $\Delta_{g}$ was found to be 1.5 meV which is similar to $\Delta$ ($\sim$ 1.8 meV) obtained from the $\rho$ data. The increase in T also suppresses the non-linearity as seen by a substantial suppression of $g_{\alpha}$. Above a few K, the transport is mainly through non-tunnelling mechanisms as increased scattering in the GB region makes the tunnelling process very weak.

\noindent
An important observation from our work is the strong effect of the magnetic field on I-V curves at low temperatures. Application of the magnetic field enhances the conductivity (both $g_{\alpha}$ and $g_0$) and suppresses the gap like feature in the $I-V$ curves.   The enhancement of $g_{\alpha}$ is very steep below 0.2 T (by nearly 2 orders) which is qualitatively similar to the LFMR and then it changes  gradually beyond that field and the change saturates beyond 14 T. The relative enhancement of the non-linear term $g_{\alpha}$ with application of field in the low field region is much more than the term $g_{0}$. For instance, between zero field and a field of 1 T the ratio $\frac{g_{\alpha}}{g_0}$ changes by 2 orders from $2\times 10^{-5}$  to $2\times 10^{-3}$ and beyond that it stays constant at this value as the field is increased. Thus at higher field, since the ratio $\frac{g_{\alpha}}{g_0}$ is field independent, the change in both occurs mainly due to barrier height change in the magnetic field as one would expect from a tunnelling process controlled by a barrier\cite{wolf}. 

\noindent
There is a rapid rise in the junction current I beyond a bias of 35 meV (see Fig. 7). This is most visible in the I-V data under high field and at low temperatures. We find that the rapid rise of the current typically occurs at a given power $P_{critical} \sim$ 130-150 nW at 0.3 K (for the data in different magnetic fields) and at somewhat higher powers as the temperature increases. We conclude that this rapid rise is a manifestation of thermal run-away which is most visible in the data taken with a magnetic field, because the junctions due to high conductance carry a high current and dissipate higher power. It is for this reason that our data analysis has been restricted to lower bias with measurement power $<$ 0.5$P_{critical}$.

\noindent
The investigation done here brings out the difference between a strongly spin polarized system like LCMO (half-metallic) and a system lacking such strong spin polarization, namely, LSCO. The absence of spin polarization in LSCO is reflected in the absence of a discernible LFMR. The grain boundary in LSCO is much less resistive, the transport shows a low value of $\Delta$ and the I-V curves are almost linear showing very weak contribution of tunnelling in case of LSCO. 

\section{CONCLUSIONS}
In summary, we have presented a comprehensive study of the electrical transport properties (including non-linear transport) of nanocrystals of two ferromagnetic perovskite oxides $\mathrm{La_{0.67}Ca_{0.33}MnO_3}$ and $\mathrm{La_{0.5}Sr_{0.5}CoO_3}$ which vary significantly in their degree of spin polarization. The studies, done in the temperature range down to 0.3 K and magnetic fields upto 14 T, allow us to investigate the effect of spin polarization in a regime where the inter-grain transport is dominated by tunnelling in Coulomb blockade  regime in such nanocrystals with sizes down to 15 nm. It is noted that this is the first time such measurements were carried out well below 1 K where the T $\ll E_C/k_B$.  The transport below 5 K was found to contain contributions of transport paths that are mostly of tunnelling type dominated by Coulomb blockade. The activation energy of transport, $\Delta$, was found to depend on the ratio $\frac{s}{d}$. From this dependence of $\Delta$ on $\frac{s}{d}$  the inverse decay length of the tunnelling wave function $\chi$, the height of the tunnelling barrier $\Phi_B$ and also their magnetic field dependences were determined. The data taken over a large magnetic field range allowed us to separate out the MR contributions at low temperatures to two distinct contributions.  At low magnetic field, the transport and the MR are dominated by the spin polarization, while at higher magnetic field the MR arises from the lowering of the tunneling barrier  by the magnetic field (due to Zeeman energy) leading to an MR that does not saturate even at 14 T. The idea of inter-grain tunnelling through the disordered grain boundary region (which acts as a barrier to tunnelling) has been validated by direct measurements of the non-linear I-V data in this temperature range. We made the important observation that at lowest temperature a gap like feature (with magnitude $\sim E_C$) shows up in the $g(V)$ versus V curve for the systems with smallest nanocrystal size. The gap closes as the field and the temperature are raised and one observes a conventional barrier type tunnelling with $g(V)\propto V^{2}$. The non-linear transport was found to be strongly dependent on the applied magnetic field.

\noindent
\section{ACKNOWLEDGEMENTS}
The authors would like to thank the Department of Science and Technology, Govt. of India for financial support in the form of a Unit (UNANST-II). One of the authors (TS) also thanks UGC, Govt. of India for a fellowship and MVK thanks CSIR, Govt. of India for a fellowship. The low temperature magnetotransport measurements were done at the Low Temperature Laboratory, UGC-DAE Consortium for Scientific Research, Indore, India. We thank Prof. P. Chaddha for allowing us to use the facility and Dr. V. Ganesan and Mr. L. S. Sharath Chandra for their help during data acquisition.

\end{document}